\documentclass{ifacconf}

\makeatletter
\newcommand\footnoteref[1]{\protected@xdef\@thefnmark{\ref{#1}}\@footnotemark}
\makeatother

\usepackage{cuted}
\usepackage{stfloats}

\usepackage{graphicx}
\graphicspath{{figures/}}

\usepackage[font=singlespacing]{caption}
\usepackage{subcaption}

\usepackage{amsmath,amssymb,amsfonts}
\usepackage{bm}
\usepackage{bbm,dsfont}
\usepackage[a]{esvect}
\usepackage{wasysym}

\allowdisplaybreaks

\DeclareMathOperator{\atanTwo}{atan2}

\let\theoremstyle\relax
\usepackage{amsthm}

\newtheorem{theorem}{Theorem}

\newtheorem{assumption}{Assumption}

\newtheorem{remark}{Remark}

\newtheorem{problem}{Problem}

\usepackage{multirow}
\usepackage{booktabs}
\usepackage{array}
\usepackage{enumerate}
\usepackage{url}
\usepackage{balance}
\usepackage{textcomp}
\usepackage{setspace}
\usepackage[dvipsnames]{xcolor}

\usepackage{natbib}

\usepackage{epstopdf}
\newcommand{\sy}[1]{{\color{black} #1}}
\newcommand{\xue}[1]{{\color{black} #1}}
\newcommand{\Xingjian}[1]{{\color{black} #1}}

\begin{document}
\begin{frontmatter}

\title{Robust Tracking of Curvature-Constrained Paths for Uncertain Dubins Systems}

\thanks{This work was supported in part by
the NASA Early Career Faculty grant 80NSSC21K0071,  ONR grant N00014-23-1-2093, and NSF grant CNS-2313814.}

\author{Xingjian Xue} \qquad  
\author{Sze Zheng Yong} 

\address{Mechanical and Industrial Engineering Department,\\  Northeastern University, Boston, MA, USA\\ (e-mail: \{xue.xing,s.yong\}@northeastern.edu)}

\begin{abstract} 
This paper presents a robust  tracking controller for  tracking curvature-constrained paths by vehicles/robots with uncertain Dubins dynamics. Although Dubins paths have been widely used in vehicular and robotic applications, robust and convergent tracking under model uncertainties remains understudied. To address this, we propose path tracking controllers based on sliding mode control, formulated in the transverse coordinate frame, which guarantee invariance and convergence of both lateral and heading errors to zero in the presence of bounded disturbances. 
Simulation results show that the proposed method reliably tracks paths despite disturbances and significantly outperforms existing methods based on sliding mode controllers. 
\end{abstract}


\end{frontmatter}

\section{Introduction}
\label{sec:introduction}

Dubins curves remain fundamental in motion planning (e.g., \cite{barth,Dolgov2008PracticalST,SHANMUGAVEL20101084,
 siedentop2015path,chipade2024withy,wiTAH,ZHANG2025120702}) 
for modern autonomous vehicles, from ground robots and warehouse Automated Guided Vehicles to Unmanned Aerial Vehicles and planetary rovers. Despite their simplicity, Dubins paths capture the curvature limits imposed by real actuators, making them highly relevant in practice. However, tracking such paths reliably under disturbances remains a challenging problem, especially when vehicle/robot dynamics deviate from idealized assumptions, e.g., due to wheel slippage (\cite{mera1,wang2008modeling}).

Path and trajectory tracking controllers for Dubins vehicles have been studied in several papers, e.g., \cite{balluchi1996path,mera1,mera2,jha2019robust}. In the ideal disturbance-free setting, \cite{balluchi1996path} proposed an elegant path tracking control law that relies only on the lateral distance to the path and the heading error as feedback. This method employs sliding mode control (SMC) (\cite{utkin1992sliding,slotine1991applied}), a nonlinear technique renowned for its robustness to matched uncertainties, strong disturbance rejection, and finite-time convergence in tracking tasks (\cite{balluchi1996path,doriacerezo2019sliding,768190}). 
Under mild curvature and smoothness assumptions, the controller guarantees convergence to the reference curvature-constrained path. However, these guarantees break down in the presence of external disturbances such as uncertain road conditions, wheel slippage, or wind. 


In contrast, \cite{mera1,mera2} proposed robust sliding mode controllers for trajectory tracking (to be differentiated from path tracking, since trajectory tracking not only tracks the deviations from the path in the transverse direction, but also the longitudinal dynamics of a reference trajectory), while effectively rejecting the disturbances. Specifically, \cite{mera1} proposed a first-order sliding-mode (FOSM) trajectory tracking  controller for  perturbed unicycle mobile robots that relied on a piecewise nonlinear sliding surface, while 
\cite{mera2} proposed an extension involving a second-order sliding mode (SOSM) controller, enforcing the convergence of both the sliding variable and its first derivative to zero. 
On the other hand, \cite{jha2019robust} addressed the noisy/uncertain setting using a robust linear-quadratic differential game (LQDG) framework for Dubins vehicle path/trajectory tracking. The nonlinear dynamics are linearized along the nominal/reference trajectory, and an LQDG controller is designed to minimize tracking error. 
However, since disturbances are only penalized through a modified cost function and the approach relies on linearization, robust trajectory tracking is not guaranteed, and performance is generally poor when there are initial state errors. 


\emph{Contributions.} In this paper, we  propose a robust path tracking controller for Dubins vehicles subject to bounded disturbances for following a curvature-constrained reference \xue{path}. In particular, our method borrows from and extends the sliding mode controller approach of \cite{balluchi1996path} that was designed for disturbance-free Dubins dynamics to ensure robustness to uncertainties such as wheel slippage, terrain effects, or wind. By formulating the problem in a transverse coordinate frame (similar to \cite{balluchi1996path}) and leveraging the disturbance rejection properties of sliding mode control, we guarantee invariance and convergence of both lateral and heading errors to zero despite disturbances. 
%
The proposed controller is validated through simulations for tracking curvature-constrained paths and benchmarked against existing robust sliding mode control tracking strategies, demonstrating improved tracking performance despite disturbances.

\section{Problem Formulation}
 \vspace{-0.1cm}
\emph{System Model.} We consider autonomous 
 mobile robots or vehicles with uncertain Dubins dynamics that are subject to disturbances (similar to \cite{mera2} but with a fixed nominal speed), as described below:
\begin{equation}\label{eq:unicycle model}
\begin{aligned}
\dot{x} &= \cos(\theta)\,(1 + d_1(t))v,\, \\
\dot{y} &= \sin(\theta)\,(1 + d_1(t))v, \\
\dot{\theta} &= (1 + d_2(t))\omega,
\end{aligned}
\end{equation}
where $(x,y)$ is the position in a global frame and $\theta$ is the heading, with the system  moving at a constant \sy{(maximum forward)} nominal speed $v=v_n$, the turn rate control input being $\omega$ and $(x_0,y_0,\theta_0)$ being the initial pose at time $t=0$. The nominal control input $\omega$ is subject to a curvature constraint given by
\begin{gather}\label{eq:input_constraint}
    \left|\frac{\omega}{v}\right|\le \frac{1}{R},
\end{gather}
where $R$ is the minimum turning radius of the vehicle/robot with Dubins dynamics in \eqref{eq:unicycle model}. Both the speed and turn rate input are subject to unknown time-varying disturbances, $d_1(t)$ and $d_2(t)$, respectively, that are assumed to be bounded, i.e., $|d_1(t)|\le \bar{d}_1 < 1$ and $|d_2(t)|\le \bar{d}_2 < 1$.
Note that in contrast to \cite{balluchi1996path}, this model from \cite{mera2} includes the consideration of bounded disturbances, where $d_1(t)$ and $d_2(t)$ are perturbations and uncertainties
related to unmodeled phenomena, e.g., caused by wheel slippage (cf. \cite{mera1,wang2008modeling}). Further, the practicality of such a model for a genuine application has been demonstrated in physical experiments in \cite{mera1,mera2}, where similar tracking controllers have been successfully deployed, although it is important to note that we aim to design path tracking/following controllers instead of trajectory tracking controllers, where the latter additionally requires the tracking of the (time-parameterized) longitudinal dynamics along the reference path.

\emph{Path Tracking Objective.} This paper aims to design a tracking controller to track a given  curvature-constrained reference path, defined as a parametrized curve
\begin{equation}
\hat{g}(s)=(\hat{x}(s),\hat{y}(s)), s \in [0,1],
\end{equation}
with $s=0$ and $s=1$ denoting the start and end of the reference path (to be distinguished from a time-parameterized reference trajectory). 

Similar to \cite{balluchi1996path}, the reference path is assumed to satisfy some constraints.\\[-0.2cm]

\begin{assumption}[Reference Path Constraints]\label{assm:ref_constraints} The following restrictions are imposed on the reference path:
\begin{enumerate}[A)]
    \item $\hat{g}(s)$ is continuously differentiable with respect to $s$, its second derivative $\hat{g}''(s)$ is allowed to have only finitely many discontinuities, and it can change its sign only a finite number of times within the interval $[0,1]$. 
    \item Let \xue{$\hat{R}(s)$ 
    be the radius of curvature of $\hat{g}(s)$,  defined as $\hat{R}(\xue{s}) = +\infty$ at inflectional points where 
$\hat{x}'(s)\hat{y}''(s) - \hat{y}'(s)\hat{x}''(s) = 0$ and otherwise},  \[
\frac{1}{\hat{R}(\xue{s})} = \frac{\hat{x}'(s)\hat{y}''(s) - \hat{y}'(s)\hat{x}''(s)}{(\hat{x}'(s)^{2} + \hat{y}'(s)^{2})^{3/2}}.
\]
The radius of curvature of $\hat{g}(s)$, $\hat{R}(\xue{s})$ is larger than or equal in magnitude to a  minimum turning radius to be determined, $\underline{R}$ (due to disturbances; cf. Theorem \ref{thm:invariance}), which is in turn larger than or equal to the minimum turning radius of the vehicle/robot, $R$, i.e.,
$$|\hat{R}(\xue{s})|\ge \underline{R} \ge R.$$
\item Consider an open neighborhood of the reference path
\[
N_\gamma = \{ n \in \mathbb{R}^2 : \exists\, s \in (0,1), \ \|n - \hat{g}(s)\| < R \}.
\]
For all $n\in N_\gamma$, there exists a unique nearest point $\hat{s} \in [0,1]$ on $\hat{g}$ such that 
\[
\|\xue{n} - \hat{g}(s)\| > \|\xue{n} - \hat{g}(\hat{s})\|, \quad \forall s \neq \hat{s}.
\]
\end{enumerate}

\end{assumption}

Note that the third restriction in Assumption \ref{assm:ref_constraints} on the  reference path may be relaxed by subdividing the path into multiple segments where each segment satisfies it.
\begin{figure}[h]  
    \centering
    \includegraphics[width=0.4\textwidth,trim={-1cm 1.2cm -1cm 2cm}]{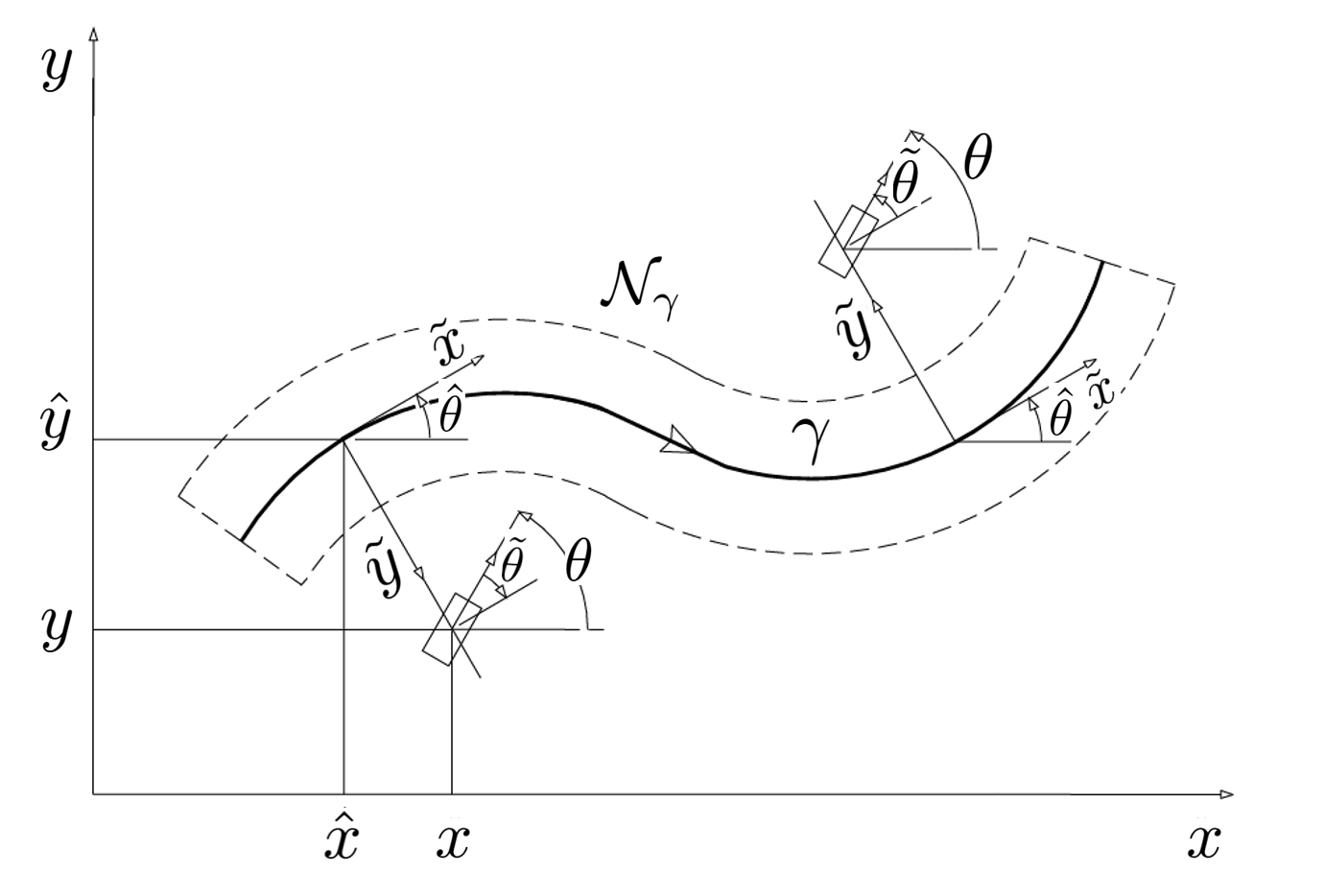}  
    \caption{Reference path and coordinates associated
 with the vehicle/robot poses 
 (similar to \cite{balluchi1996path}).}
 \vspace{-0.6cm}
    \label{fig2}
\end{figure}

Next, as in \cite{balluchi1996path}, we consider the transverse coordinate frame (also known as the Frenet–Serret frame; cf. Fig. \ref{fig2} for an illustration) to simplify the design and analysis of the robust path tracking controller. To obtain this frame, let $\hat{\theta}(s)$ denote the orientation of the tangent to the reference path with respect to the $x$-axis of the world frame,
\[
\hat{\theta}(s) = \atanTwo \bigl(\hat{y}'(s), \hat{x}'(s)\bigr),
\]
and define the local Frenet coordinate as $(\tilde{x}, \tilde{y}, \tilde{\theta})$, where  $\tilde{x}$ is the longitudinal error of the vehicle/robot relative to $\hat{g}({s})$, $\tilde{y}$ is the lateral error, and  $\tilde{\theta}$ is the heading error: 
\begin{align}\label{eq:frenet error}
\begin{array}{rl}
\tilde{x}(x,y,s)\!\! &= (x - \hat{x}(s)) \cos(\hat{\theta}(s))  + (y - \hat{y}(s)) \sin(\hat{\theta}(s)), \\
\tilde{y}(x,y,s)\!\! &= \operatorname{sign}(\hat{R}(\hat{s})) 
[ (y - \hat{y}(s)) \cos(\hat{\theta}(s))\\ 
&\quad - (x - \hat{x}(s)) \sin(\hat{\theta}(s))], \\
\tilde{\theta}(\theta,s)\!\! &= \operatorname{sign}(\hat{R}(\hat{s})) (\theta - \hat{\theta}(s)). 
\end{array}
\end{align}

By Assumption \ref{assm:ref_constraints}-B, we can associate each position $(x(t),y(t))$ at time $t$ \xue{with} a unique point $\hat{s}$ on the reference path $\hat{g}$ that is closest to $(x(t),y(t))$. This can be implicitly determined via \eqref{eq:frenet error} by setting 
\begin{gather}\label{eq:implicit}
    \tilde{x}(x(t),y(t),\hat{s}) = 0,
\end{gather}
such that the focus lies on performing path tracking (in the transverse coordinates only), which should be differentiated from  trajectory tracking, e.g., in \cite{mera1,mera2} with sliding mode controllers, where the time-parameterized longitudinal trajectory also needs to be tracked.  
Then, by differentiating \eqref{eq:frenet error} and \eqref{eq:implicit} and applying the implicit function theorem, we can obtain the transverse/Frenet frame dynamics for the lateral and heading errors, which is applicable almost everywhere, 
\begin{equation}
\begin{aligned}
\dot{\tilde{y}} &= \sin(\tilde{\theta})\,(1 + d_1(t))v,   \\
\dot{\tilde{\theta}} &= 
-\frac{\cos(\tilde{\theta})}{|\hat{R}(\hat{s})| - \tilde{y}} \, (1 + d_1(t))v 
+ \operatorname{sign}(\hat{R}(\hat{s}))\,(1 + d_2(t))\omega,
\end{aligned}\label{eq:transverse}
\end{equation}
where $\hat{s}(t)$ is determined as the (unique) point on the reference path $\hat{g}$ that is closest to the actual position $(x(t),y(t))$.

In the state space construction, the state vector is defined as $[x, y, \theta]^{T}$. All the Frenet state quantities can be obtained \xue{from} \eqref{eq:frenet error} and the state vector to construct the sliding surface and signal input. 

Using this new transverse coordinate frame, the goal of our paper can be 
stated as follows:\\[-0.2cm]
\begin{problem}\label{prob:1}
Design a  path tracking control law $\omega(\tilde y(t),\tilde\theta(t)$, $\operatorname{sign}(\hat{R}(\hat{s}(t)))$ satisfying \eqref{eq:input_constraint} such that  for any initial pose
$(x_0, y_0, \theta_0)$ of the robot in a suitable neighborhood
of the reference path, denoted as the invariant set $\mathcal{S}$,
\begin{enumerate}[1)] 
\item the set $\mathcal{S} \subseteq [-R,R] \times [-\frac{\pi}{2},\frac{\pi}{2}]$ remains forward invariant, i.e., such that $(\tilde{y}(t),\tilde{\theta}(t)) \in \mathcal{S}$ for all $t\ge 0$ and 
\item $(\tilde{y},\tilde{\theta})$ converges to zero irrespective of the bounded unknown disturbances.
\end{enumerate}
 
\end{problem}

 

\section{Main results}\label{sec:main}
Inspired by the sliding mode controller  approach in \cite{balluchi1996path} for disturbance-free Dubins robots, we propose to solve Problem \ref{prob:1} using a similar sliding mode controller given by 
\begin{equation}\label{eq:controller}
\omega = \operatorname{sign}(\hat{R}(\hat{s}))  \operatorname{sign}(\sigma(\tilde{y},\tilde{\theta}))  \tfrac{v}{R},
\end{equation}
where $\sigma$ is the sliding variable defined as
\begin{equation}\label{eq:sliding manifold}
\sigma(\tilde{y}, \tilde{\theta}) = -\tfrac{\tilde{y}(1-q)}{R} - \operatorname{sign}(\tilde{\theta})(1 - \cos(\tilde{\theta}))
\end{equation}
and $q \in [0,1)$ is a to-be-determined scalar (cf. Theorem \ref{theorem 2}) for ensuring robustness \xue{to disturbances}. Note that when $q=0$, the sliding variable reduces to the one considered for the disturbance-free setting in \cite{balluchi1996path}.

\subsection{Forward Invariance of Lateral and Heading Errors}
First, to address Problem \ref{prob:1}.1, we construct a forward invariant set $\mathcal{S}\subseteq [-R,R] \times [-\frac{\pi}{2},\frac{\pi}{2}]$ with the proposed feedback law in \eqref{eq:controller}. All proofs are provided in the Appendix for the sake of increased readability.\\

\begin{figure}[t]  
    \centering
    \includegraphics[width=0.35\textwidth,trim={0.1cm, 0.5cm, 0.2cm, 0cm},clip]{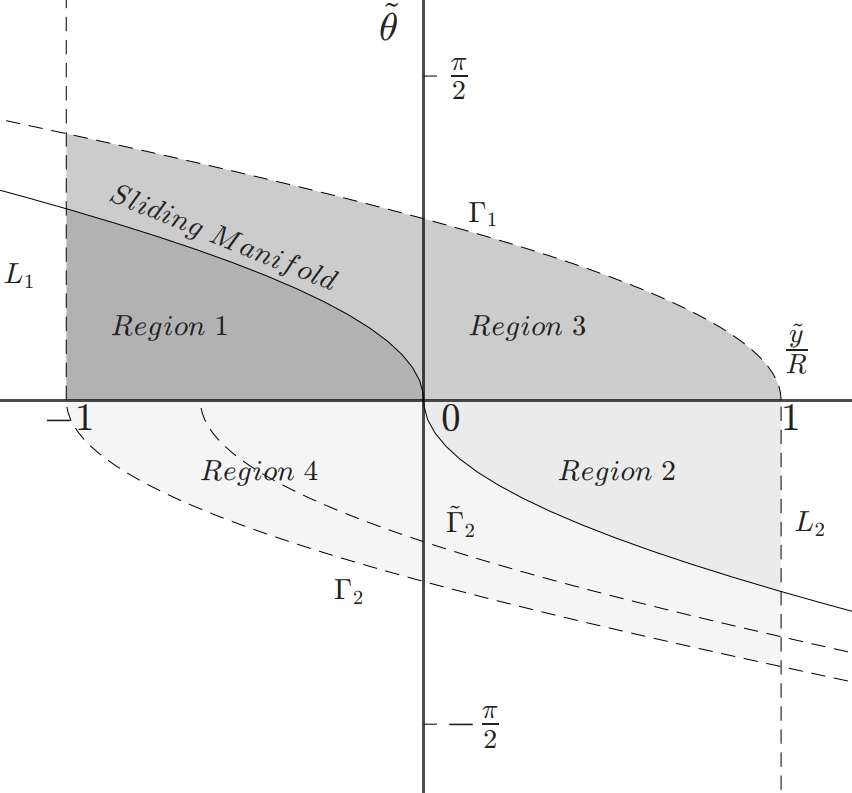}  
    \caption{Illustration of the invariant set $\mathcal{S}$ (gray regions) and the sliding surface/manifold $\sigma=0$.}
    \label{fig:invariantset}
\end{figure}

\begin{theorem}[Robust Invariance of Tracking Errors]\label{thm:invariance}
Given a robot with uncertain Dubins dynamics in~\eqref{eq:unicycle model} and a reference path satisfying Assumption \ref{assm:ref_constraints} with 
 \begin{equation}\label{eq:R_nominal_minimum}
\begin{array}{c}|\hat{R}(\hat{s})|\geq \underline{R} :=\frac{2R}{1-p} - \big(\tfrac{\tilde{y}}{R}\big)_{d}R,\end{array}
\end{equation}
\xue{where $(\tfrac{\tilde{y}}{R})_d \leq 1$ is the intercept of the robust invariant set with the $\tfrac{\tilde y}{R}$ axis
and $p \in [0,1)$ is a tuning parameter.}
Let the feedback controller be given by ~\eqref{eq:controller} with the sliding variable~\eqref{eq:sliding manifold}. Then, the set $\mathcal{S}$ (cf. Fig. \ref{fig:invariantset}) of lateral and heading errors defined by
\begin{gather}
\begin{array}{r}
    \mathcal{S}:=\{(\tilde{y},\tilde{\theta}) \in [-R,R] \times [-\frac{\pi}{2},\frac{\pi}{2}]  \mid L_1(\tilde{y},\tilde{\theta})\ge 0, \\ L_2(\tilde{y},\tilde{\theta})\ge 0, \Gamma_1(\tilde{y},\tilde{\theta})\ge0, \Gamma_2(\tilde{y},\tilde{\theta})\ge 0\},
    \end{array}
\end{gather}
with 
\begin{gather}
    L_{1}(\tilde{y}, \tilde{\theta}) := \tfrac{\tilde{y}}{R}+ (\tfrac{\tilde{y}}{R})_{d}, \label{eq:L1} \\
    L_{2}(\tilde{y}, \tilde{\theta}) := (\tfrac{\tilde{y}}{R})_{d}- \tfrac{\tilde{y}}{R}, \label{eq:L2} \\
    \Gamma_{1}(\tilde{y}, \tilde{\theta}) := -2 +(\tfrac{\tilde{y}}{R})_{d}(1\!-\!p)- \tfrac{\tilde{y}}{R}(1 \!-\! p) + 2 \cos \tilde{\theta}, \label{eq:gamma1}\\
    \Gamma_{2}(\tilde{y}, \tilde{\theta}) := -2 +(\tfrac{\tilde{y}}{R})_{d}(1\!-\!p)+ \tfrac{\tilde{y}}{R}(1 \!-\! p) + 2 \cos \tilde{\theta}, \label{eq:gamma2}
\end{gather}
where $p$ 
satisfies
\begin{gather}\label{eq:p_range}
1 > p \geq 1 - \frac{1 - \bar{d}_2}{1 + \bar{d}_1}\ge 0,
\end{gather}
is forward invariant, i.e., $(\tilde{y}(0),\tilde{\theta}(0))\in \mathcal{S}$ at time $t=0$ implies that  $(\tilde{y}(t),\tilde{\theta}(t))\in \mathcal{S}$ for all times $t\ge0$. In other words, the lateral and heading tracking errors $(\tilde{y}(t),\tilde{\theta}(t))\in \mathcal{S}$ remain bounded at all times.
\end{theorem}

Note that in the above, the choice of $p$ is dependent on the disturbance bounds $\bar{d}_1$ and $\bar{d}_2$; hence, the minimum curvature radius of the reference path $\underline{R}$ in \eqref{eq:R_nominal_minimum} must also increase to ensure robustness \xue{to disturbances}. In the special case when there is no disturbance, i.e., $\bar{d}_1=\bar{d}_2=0$, $p$ can be chosen to be 0 and since $(\tfrac{\tilde{y}}{R}) \le 1$, \eqref{eq:R_nominal_minimum} becomes $|\hat{R}(\hat{s})| \ge R$, which was one of the assumptions in \cite{balluchi1996path}.
This $\underline{R}$ can be shown to be always larger than the minimum turning radius $R$ 
and is increasing as the disturbance magnitudes increase. \\

\begin{remark}\label{rem1} To ensure broader applicability of the approach to various initial tracking errors, 
we propose to construct the \xue{maximal/largest} $\mathcal{S}$ 
by appropriately selecting  $p$ and $(\tfrac{\tilde{y}}{R})_d$. Specifically, since, by Assumption \ref{assm:ref_constraints}-C, the robot must be in the neighborhood of the reference path with distances to the reference path being smaller than $R$, the maximal $\mathcal{S}$ can be obtained with $(\tfrac{\tilde{y}}{R})_{d} = 1$. Further, from \eqref{eq:gamma1} and \eqref{eq:gamma2}, we observe that the parameter $p$ has the effect \xue{of compressing}  $\mathcal{S}$ in the vertical direction, so the maximal $\mathcal{S}$ can be obtained with 
$p = 1 - \frac{1 - \bar{d}_2}{1 + \bar{d}_1}$.
\end{remark}

\subsection{Sliding Mode Control}
Next, we address Problem \ref{prob:1}.2 by proving that the proposed feedback control law in \eqref{eq:controller} with the sliding variable in \eqref{eq:sliding manifold} leads to the tracking errors converging to zero.\\ 

\begin{theorem}[Convergence of Tracking Errors]\label{theorem 2}
Given a robot with uncertain Dubins dynamics in~\eqref{eq:unicycle model} and a reference path satisfying Assumption \ref{assm:ref_constraints} with $\underline{R}$ satisfying \eqref{eq:R_nominal_minimum}, employing the  feedback controller given by ~\eqref{eq:controller} with the sliding variable~\eqref{eq:sliding manifold} and $q $ that satisfies
\begin{equation}\label{constraint for q}
0 \le 1- \frac{(1 - \bar{d}_2)}{(1 + \bar{d}_1)} \leq q \leq \frac{(1 - \bar{d}_2)}{(1 + \bar{d}_1)}\le 1,
\end{equation}
results in the lateral and heading tracking errors converging to zero, i.e., $\lim_{t\to \infty} (\tilde{y}(t),\tilde{\theta}(t))=(0,0)$.

\end{theorem}
Furthermore, note that the restriction in \eqref{constraint for q} induces a limitation on the disturbance bounds $\bar{d}_1$ and $\bar{d}_2$ that can be handled with the proposed approach. Specifically, from the left and right of \eqref{constraint for q}, we can deduce that the disturbance bounds $\bar{d}_1$ and $\bar{d}_2$ that can be rejected with our proposed approach must satisfy the following constraint:
\begin{equation}\label{additional constraint for q}
(1-\bar{d}_2) \geq 0.5 (1+\bar{d}_1).
\end{equation}


\begin{remark}
Although the proposed controller in \eqref{eq:controller} can, in theory, entirely reject all disturbances and perfectly track the reference path, its practical application is limited since the discontinuous nature of the sliding mode control law leads to undesirable input chattering that may cause damage to the hardware. Hence, as in \cite{slotine1991applied}, 
\sy{the controller in \eqref{eq:controller} can be modified} with a boundary layer to eliminate input chattering: 
\begin{equation}\label{eq:controller_mod}
\omega = 
\operatorname{sign}(\hat{R}(\hat{s}))  \operatorname{sat}_\phi(\sigma(\tilde{y},\tilde{\theta}))  \frac{v}{R},
\end{equation}
where the saturation function $\operatorname{sat}_\phi$ is defined as
\begin{equation}\label{eq:sat}
\operatorname{sat}_\phi(\sigma(\tilde{y},\tilde{\theta}))=\begin{cases}
    \frac{\sigma(\tilde{y},\tilde{\theta})}{\phi}, & \text{if } |\sigma(\tilde{y},\tilde{\theta})| \le  \phi,\\
    \operatorname{sign}(\sigma(\tilde{y},\tilde{\theta})), & \text{otherwise},
\end{cases}
\end{equation}
with the boundary layer width $\phi$ as a tuning parameter.

Note that the introduction of the boundary layer results in a loss of the guarantee of 
convergence of the tracking errors to zero, but instead, we observe that the tracking errors become ultimately bounded within a (minimal) invariant set, \xue{i.e., with small bounded tracking errors.} A rigorous proof of this observation is omitted due to space limitations and will be addressed in future work.
\end{remark}

\begin{figure*}[t]
    \vspace{-1.5cm}
    \centering 
    \begin{subfigure}[h]{0.24\linewidth}
    \includegraphics[width=1\linewidth,trim={5cm 5cm 5cm 5cm},clip]{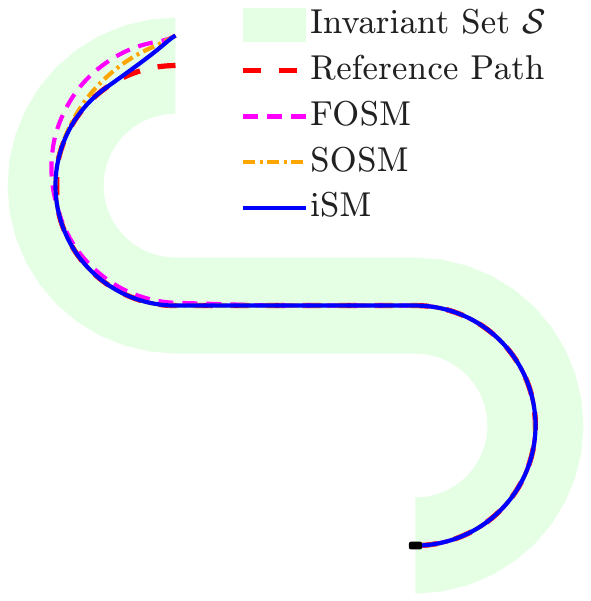}
        \label{fig:path_1}\vspace{-1.8cm}
    \end{subfigure}
    \begin{subfigure}[h]{0.24\linewidth}
    \includegraphics[width=1\linewidth,trim={5cm 5cm 5cm 5cm},clip]{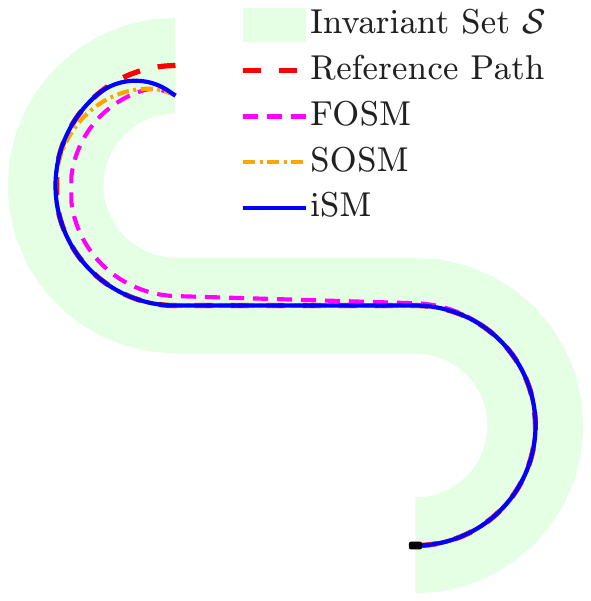}
        \label{fig:path_2} \vspace{-1.8cm}
    \end{subfigure}	
    \begin{subfigure}[h]{0.24\linewidth}
    \includegraphics[width=1\linewidth,trim={5cm 5cm 5cm 5cm},clip]{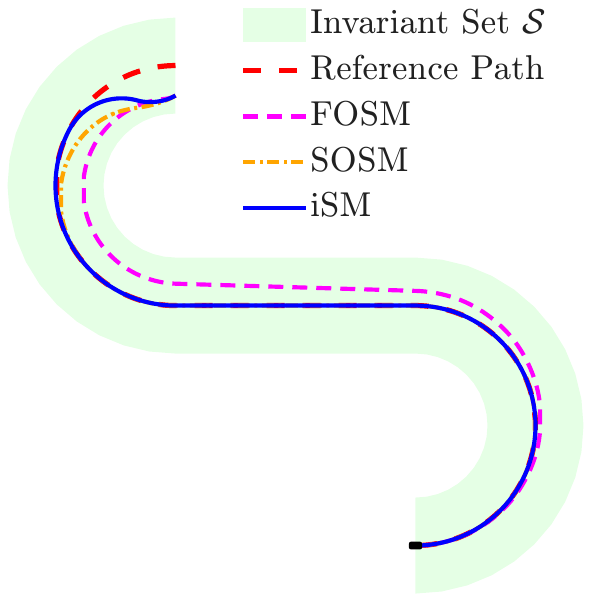}
        \label{fig:path_3} \vspace{-1.8cm}
    \end{subfigure}	
    \begin{subfigure}[h]{0.24\linewidth}
    \includegraphics[width=1\linewidth,trim={5cm 5cm 5cm 5cm},clip]{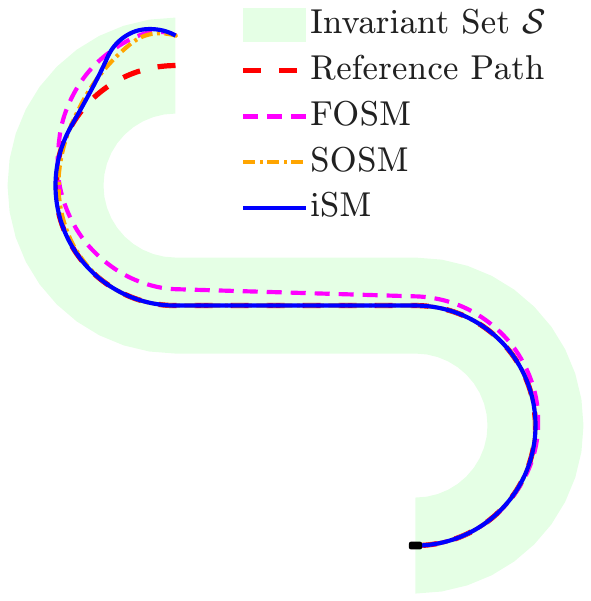}
        \label{fig:path_4}\vspace{-1.8cm}
    \end{subfigure}
    \vspace{-1cm}
    
    \begin{subfigure}[h]{0.22\linewidth}
    \includegraphics[width=1\linewidth,trim={4.8cm 4cm 5cm 4.5cm},clip]{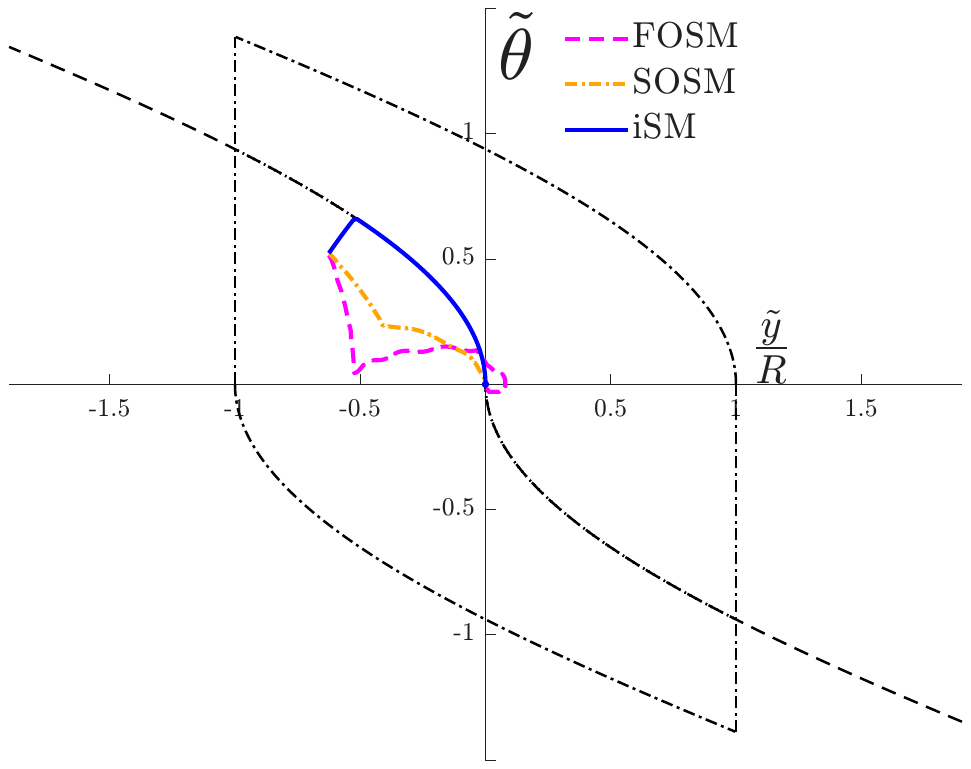}
        \label{fig:path_5}\vspace{-1.8cm}
    \end{subfigure}	
    \hspace{0.01\linewidth}
    \begin{subfigure}[h]{0.22\linewidth}
    \includegraphics[width=1\linewidth,trim={4.8cm 4cm 5cm 4.5cm},clip]{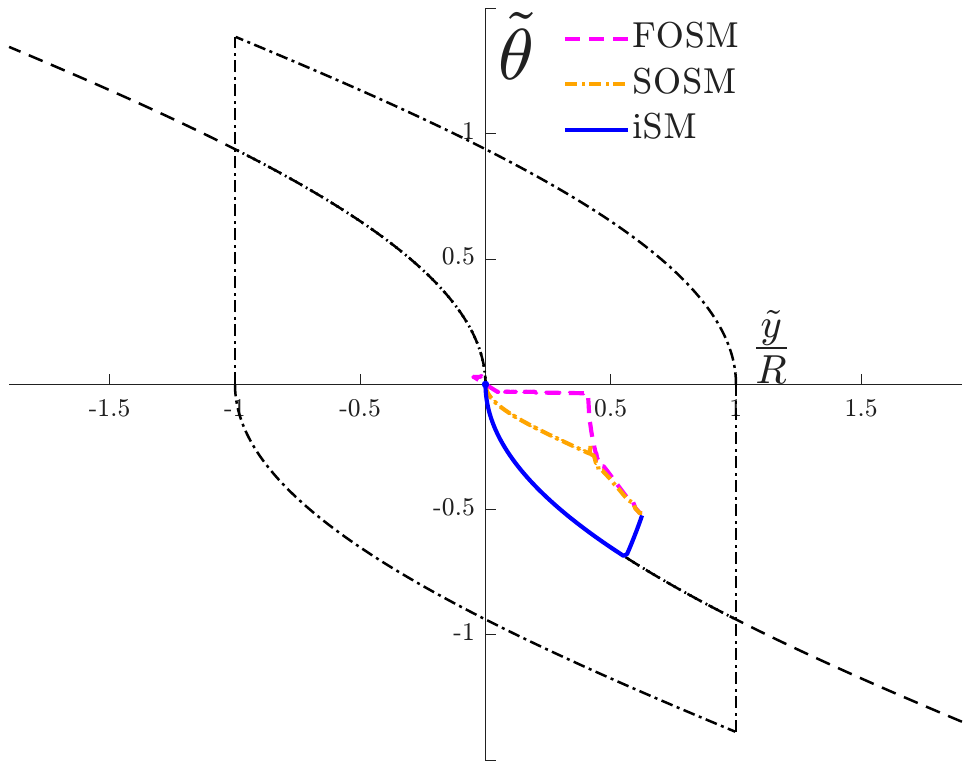}
        \label{fig:path_6} \vspace{-1.8cm}
    \end{subfigure}	
    \hspace{0.01\linewidth}
    \begin{subfigure}[h]{0.22\linewidth}
    \includegraphics[width=1\linewidth,trim={4.8cm 4cm 5cm 4.5cm},clip]{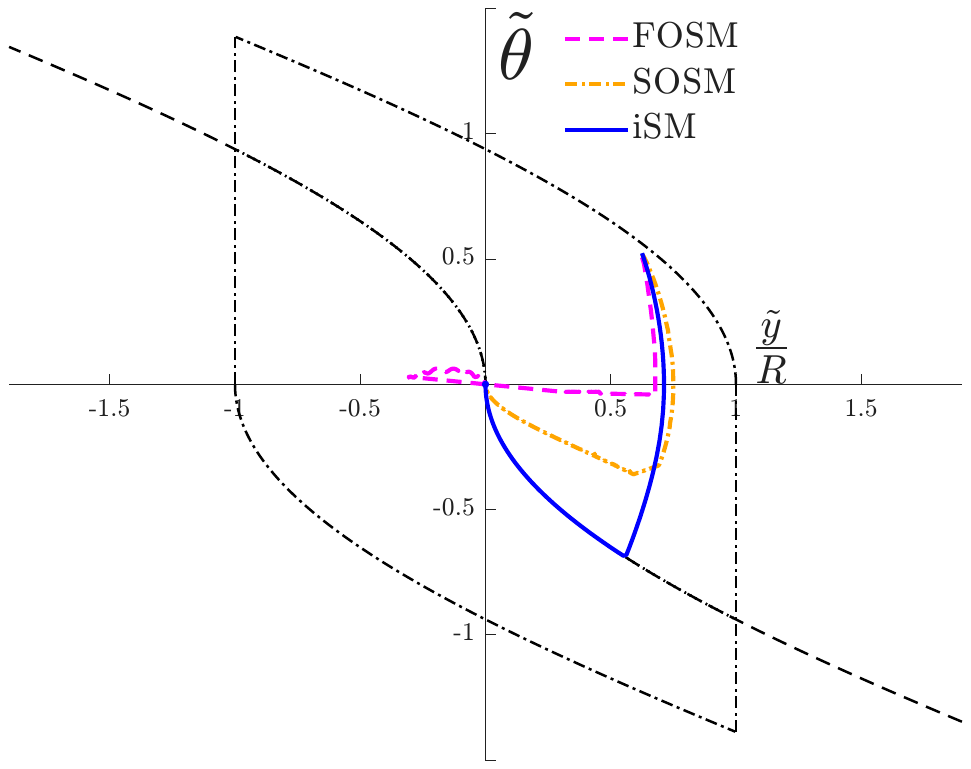}
        \label{fig:path_6} \vspace{-1.8cm}
    \end{subfigure}	
    \hspace{0.01\linewidth}
    \begin{subfigure}[h]{0.22\linewidth}
    \includegraphics[width=1\linewidth,trim={4.8cm 4cm 5cm 4.5cm},clip]{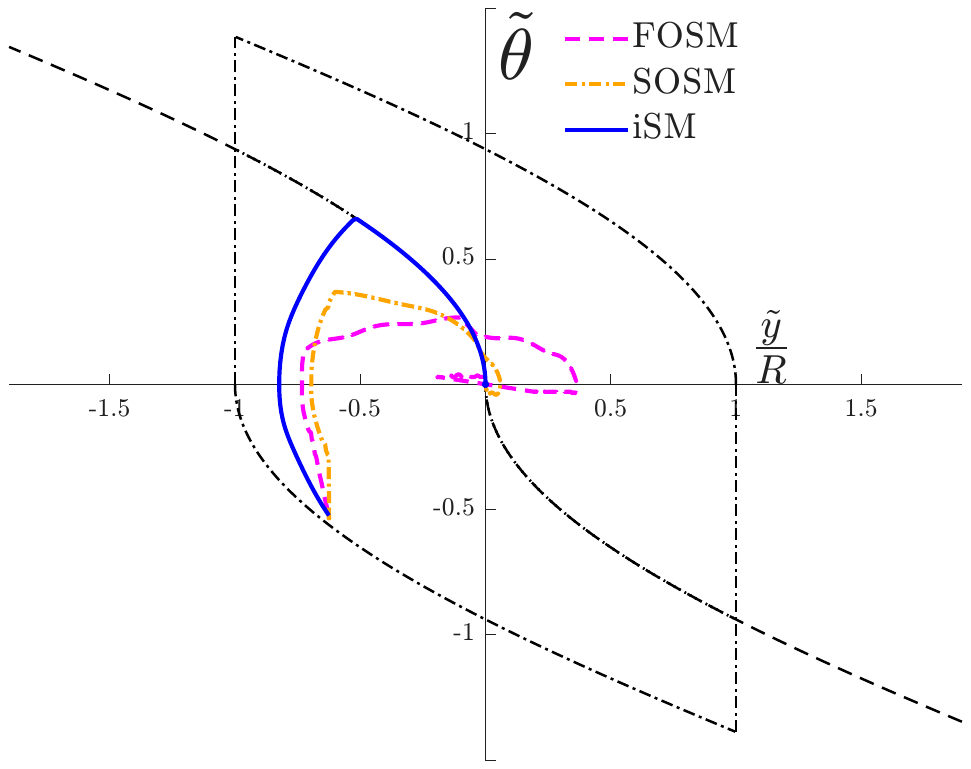}
        \label{fig:path_6} \vspace{-1.8cm}
    \end{subfigure}	
    \vspace{-1.2cm}
    
    \begin{subfigure}[h]{0.245\linewidth}
    \includegraphics[width=1\linewidth,trim={0cm 5cm 2cm 1cm},clip]{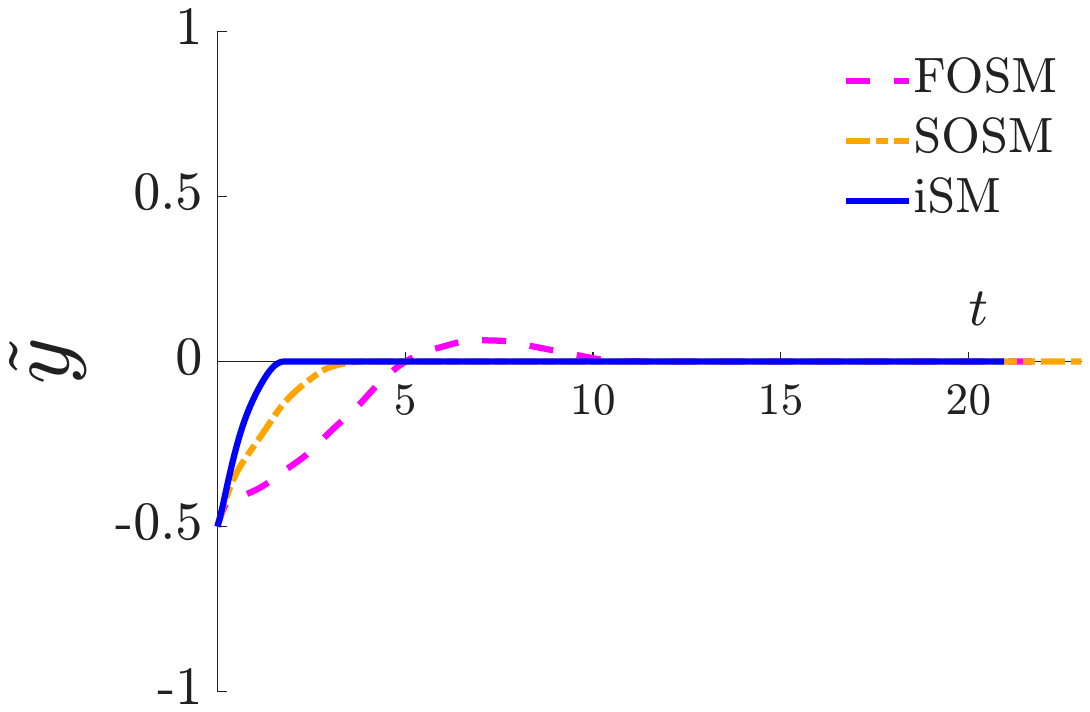}
        \label{fig:path_6} \vspace{-1.2cm}
        \caption{$\tilde y_0 = -0.5, \tilde{\theta}_0 = 30^\circ$}
    \end{subfigure}	
    \begin{subfigure}[h]{0.245\linewidth}
    \includegraphics[width=1\linewidth,trim={0cm 5cm 2cm 1cm},clip]{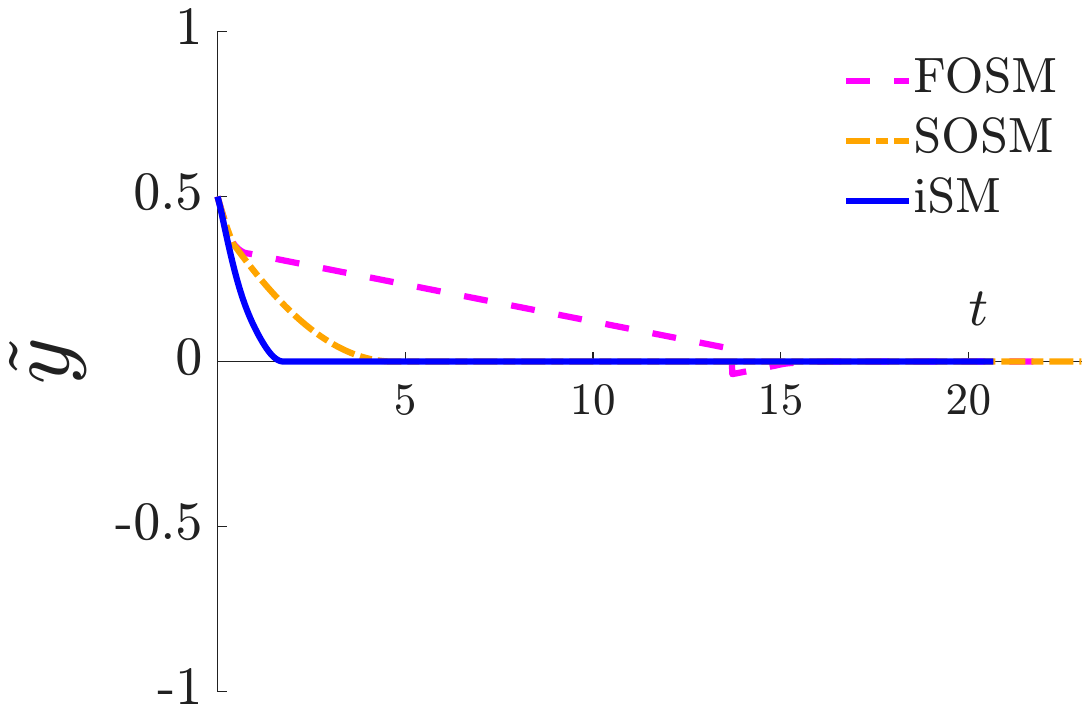}
        \label{fig:path_6} \vspace{-1.2cm}
        \caption{$\tilde y_0 = 0.5, \tilde{\theta}_0 = -30^\circ$}
    \end{subfigure}	
    \begin{subfigure}[h]{0.245\linewidth}
    \includegraphics[width=1\linewidth,trim={0cm 5cm 2cm 1cm},clip]{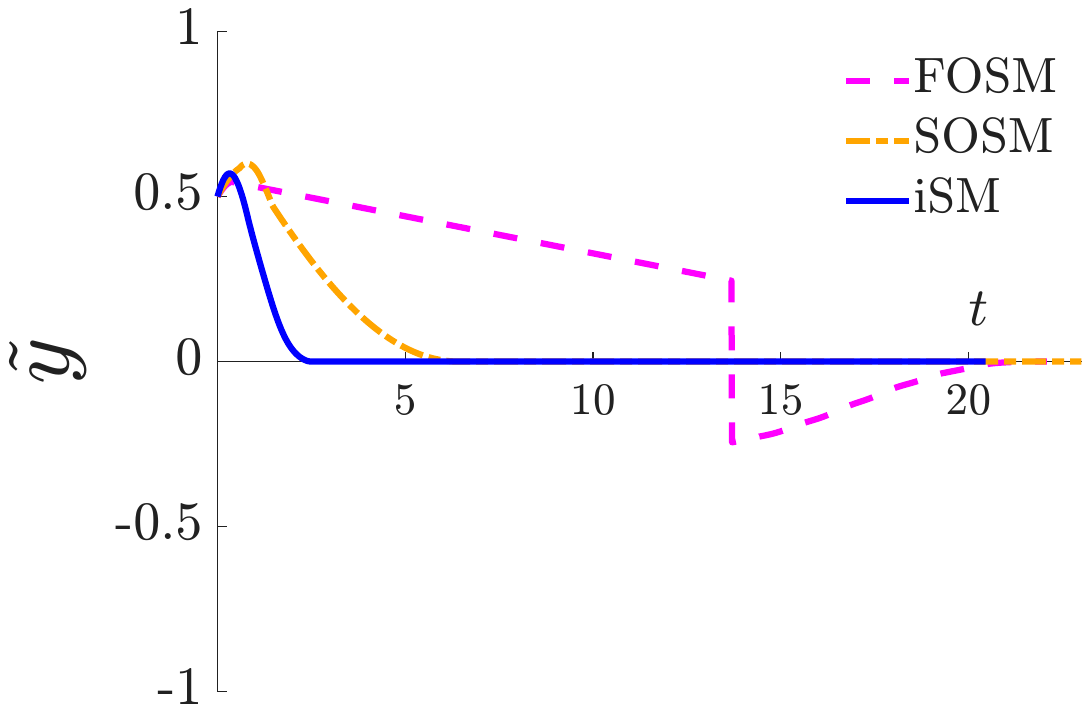}
        \label{fig:path_6} \vspace{-1.2cm}
        \caption{$\tilde y_0 = 0.5, \tilde{\theta}_0 = 30^\circ$}
    \end{subfigure}	
    \begin{subfigure}[h]{0.245\linewidth}
    \includegraphics[width=1\linewidth,trim={0cm 5cm 2cm 1cm},clip]{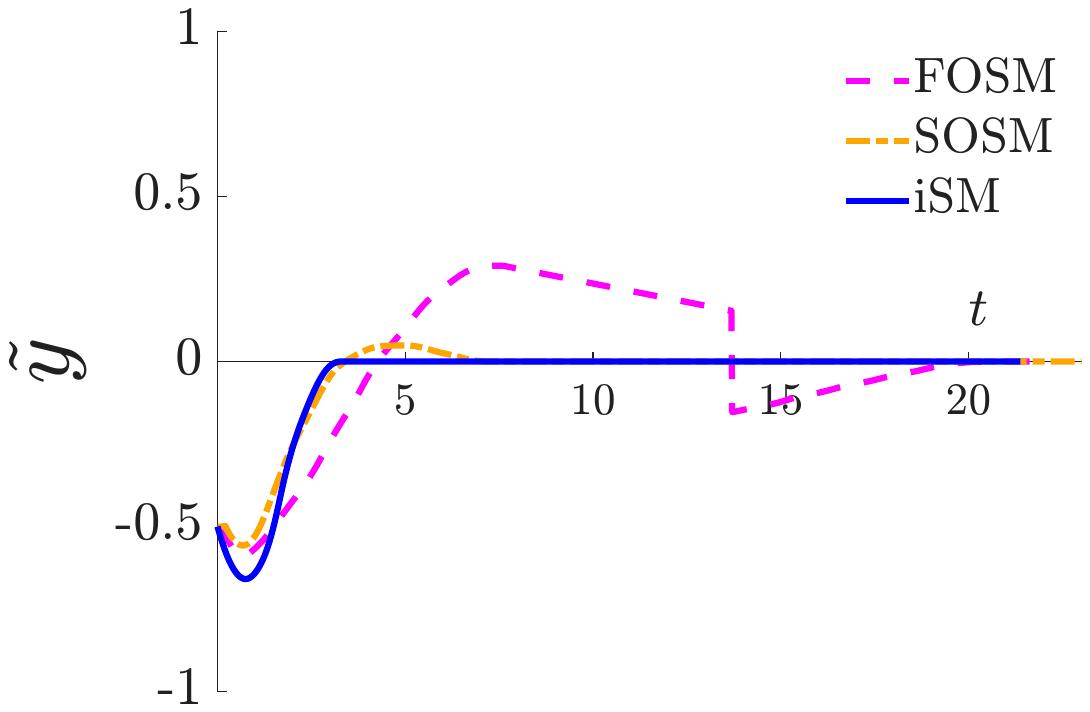}
        \label{fig:path_6} \vspace{-1.2cm}
        \caption{$\tilde y_0 = -0.5, \tilde{\theta}_0 = -30^\circ$}
    \end{subfigure}	
    \vspace{0cm}
    
    
    \caption{A comparison of FOSM (\cite{mera1}), SOSM (\cite{mera2}) and our proposed invariant sliding mode controller (iSM) with  \eqref{eq:controller}. 
    The first row depicts the path tracking performance of all three methods as well as the invariant set $\mathcal{S}$.  
    The second row provides the phase portrait of the lateral and heading error states during tracking (within $\mathcal{S}$). 
    The third row shows the lateral error trajectories of the various approaches.}
    \label{fig1}
\end{figure*}

\vspace{-0.1cm}
\section{Simulation}
 \vspace{-0.2cm}
Next, we investigate the performance of the proposed 
path tracking approach for tracking a curvature-constrained 
reference path (adapted from \cite{balluchi1996path}): 
\begin{align*}
\hat{x}(s) &=
\begin{cases}
-2\sin(4\pi s), & 0 \leq s < 0.25, \\
8(s-0.25), & 0.25 \leq s < 0.75, \\
4 + 2\sin\bigl(4\pi(s-0.75)\bigr), & 0.75 \leq s \leq 1,
\end{cases}
\\
\hat{y}(s)& =
\begin{cases}
2+2\cos(4\pi s), & 0 \leq s < 0.25, \\
0, & 0.25 \leq s < 0.75, \\
-2 + 2\cos\bigl(4\pi(s-0.75)\bigr), & 0.75 \leq s \leq 1.
\end{cases}
\end{align*}
The initial error states $[\tilde y_0, \tilde \theta_0]$ were 
chosen to be within each of the four quadrants of Fig. \ref{fig:invariantset}: $[-0.5,30^\circ]$, $[0.5, -30^\circ]$, $[0.5, 30^\circ]$, and $[-0.5, -30^\circ]$. The minimum turning radius $R$ is set to 0.8, \sy{the nominal velocity to $v_n=0.8$, and the disturbance bounds to $\bar{d_1}= \bar{d_2}= 0.1$.} 
To demonstrate the effectiveness of our 
approach, we 
compared our proposed invariant sliding mode (iSM) \emph{path} tracking  approach \sy{in \eqref{eq:controller}} with the first-order sliding mode (FOSM) and second-order sliding mode (SOSM) \emph{trajectory} tracking controllers in~\cite{mera1, mera2}, respectively, using the same initial states. \sy{In our iSM approach, $p, q$ were chosen as $0.182$ and $0.59$ respectively, while} controller parameters \sy{for FOSM and SOSM} were directly taken from \cite{mera1, mera2}, \xue{except that the tuning parameter $\beta$ was set to $0$ to remove the smooth approximation \sy{that would render their convergence guarantees invalid}. Further, the reference velocities $v_d$ for FOSM and SOSM were chosen as 0.75 and 0.72, respectively, with $\omega_d=\tfrac{v_d}{\Xingjian{\hat{R}(s)}}$ to enable error convergence by the end of the trajectory with the maximum forward velocity of 0.8.}  

From Fig.~\ref{fig1}, we observe that our proposed iSM approach \sy{achieved the fastest error convergence across all four initial states, followed by SOSM, \xue{while} FOSM \xue{is} the slowest}. 
\sy{This is because FOSM and SOSM face a trade-off between error convergence rate and forward speed, with parameters tuned to prioritize speed, whereas iSM achieves maximal forward speed without affecting error convergence.} 

Overall, our 
path tracking controller delivered better performance than
state-of-the-art trajectory tracking approaches (FOSM and SOSM) in terms of \sy{faster error convergence and forward speed.} 
Moreover, our approach is simpler to tune since it involves fewer tuning parameters ($p, q$) than the FOSM and SOSM methods,  
which can have a significant impact on 
performance. 
On the other hand, our path tracking approach may require some modifications if the initial state is far from the reference path such that the transverse coordinates are undetermined, but this can often be overcome by extrapolating the path, if needed.

 \vspace{-0.05cm}
\section{Conclusions} \label{sec:conclusions}
 \vspace{-0.15cm}
In this paper, we proposed a robust path tracking controller for a curvature-constrained reference path using sliding mode control. 
By carefully designing the sliding manifold,  
we established forward invariance of the tracking error dynamics and ensured convergence of the errors to zero despite the presence 
of bounded disturbances. 
The theoretical analysis was supported by simulation studies on tracking a composite reference 
path, where our proposed controller outperformed state-of-the-art approaches for robust trajectory tracking. Future work will rigorously characterize the ultimate bound of the tracking errors in the presence of the boundary layer for eliminating input chattering, as well as extend our results to vehicles/robots with more general uncertain 
dynamics.

 \vspace{-0.05cm}
\section{Appendix}
 \vspace{-0.2cm}
\subsection{Proof of Theorem 1}
 \vspace{-0.1cm}
    We will show that the vector fields at the four boundaries of the set $\mathcal{S}$ (cf. Fig. \ref{fig:invariantset}) point inwards to leverage Nagumo's invariance principle:

    \emph{Boundary 1: $L_1 = 0$ (with $0 \le \tilde{\theta}\le \frac{\pi}{2}$).} Differentiating $L_1$ with respect to $t$, we obtain 
    $$\dot L_{1}(\tilde{y}, \tilde{\theta}) =\frac{\sin\tilde{\theta}(1 + d_1(t))v}{R}.$$
    Since $\sin \tilde{\theta}\ge 0$ and $|d_1(t)|\le \bar{d}_1 < 1$, we have $\dot{L}_1\ge 0$, which means that $L_1$ remains non-negative (``points inwards'').
    
    \emph{Boundary 2: $L_2 = 0$ (with $-\frac{\pi}{2} \le \tilde{\theta} \le 0$).}
    Similarly,  differentiating $L_2$ w.r.t. $t$,
    $$\dot L_{2}(\tilde{y}, \tilde{\theta}) =-\frac{\sin\tilde{\theta}(1 + d_1(t))v}{R} $$ and $L_2$ remains non-negative since $\dot{L}_2\ge 0$ with $\sin \tilde{\theta}\le 0$ and $|d_1(t)|\le \bar{d}_1 < 1$. 
    
    \emph{Boundary 3: $\Gamma_1 = 0$ (with $0 \le \tilde{\theta} \le \frac{\pi}{2}, \sigma \le 0$).}
    Next, differentiating $\Gamma_1$ w.r.t. $t$, we have
\begin{align*}
\begin{array}{rl}
\dot{\Gamma}_{1} =& \sin(\tilde{\theta}) \big(
    \frac{(p - 1)(1 + d_1(t))v}{R}
    + \frac{2 \cos(\tilde{\theta})(1 + d_1(t))v}{|\hat{R}(\hat{s})| - \tilde{y}} 
    \\ 
    &- 2\operatorname{sign}(\sigma)\frac{\, (1 + d_2(t))v}{R} 
\big)\\
\ge & \sin(\tilde{\theta})( \frac{(p - 1)(1 + d_1(t))v+2(1 + d_2(t))v}{R}) \ge 0, \end{array} 
\end{align*}

where the first inequality is obtained since $\sin \tilde{\theta}\ge 0$, $\cos \tilde{\theta}\ge 0$, $\sigma \le 0$, $0\le \xue{p} <1$, $|\hat{R}|> \tilde{y}$, $|d_1(t)|\le \bar{d}_1 < 1$, and $\operatorname{sign}(\hat{R}(\hat{s})\xue{)}\in \{-1,0,1\}$, while the second is obtained if 
\begin{gather}\label{eq:p_range_for_gamma1}
p \geq \max_{\substack{|d_1(t)|\le \bar{d}_1,\\|d_2(t)|\le \bar{d}_2}}(1- \frac{2(1 + d_2(t))}{(1 + d_1(t))}) = 
1- \frac{2(1 - \bar{d}_2)}{(1 + \bar{d}_1)}.
\end{gather}
    \emph{Boundary 4: $\Gamma_2 = 0$ (with $-\frac{\pi}{2} \le \tilde{\theta} \le 0, \sigma \ge 0$).}
    Differentiating $\Gamma_2$ w.r.t. $t$, we obtain
    \begin{align}\label{eq:gamma2_derivative}
\begin{array}{rl}\dot{\Gamma}_{2} = &\sin(\tilde{\theta})( \frac{(1-p)(1 + d_1\xue{(t)})v}{R}
+ 
\frac{\Xingjian{2\cos(\tilde{\theta} )}(1 + d_1\xue{(t)})v}{|\hat{R}(\hat{s})| - \tilde{y}}
\\&-2\operatorname{sign}(\sigma)\frac{(1 + d_2\xue{(t)})v}{R}).
\end{array}
\end{align}
Unlike for Boundary 3, we 
 only prove that $\dot{\Gamma}_2 \ge 0$ in the vicinity of 
 Boundary 4 (as required by Nagumo's invariance principle), and specifically, when $0 \le \Gamma_2 \le \tilde{\Gamma}_2$ with
\begin{equation}\label{eq:gamma2_supplement_function}
\begin{array}{c}\tilde \Gamma_{2} := -2 +(\tfrac{\tilde{y}}{R})_{d}(1-p)+ \frac{|\hat{R}|}{R}(1 - p).\end{array}
\end{equation}
Then, when $\Gamma_2 \le \tilde{\Gamma}_2$, from \eqref{eq:gamma2} and \eqref{eq:gamma2_supplement_function}, we have
$$\begin{array}{c}2 \cos \tilde{\theta} \le \frac{|\hat{R}|-\tilde{y}}{R}(1 - p)\Rightarrow \frac{2 \cos \tilde{\theta}}{|\hat{R}|-\tilde{y}}=\frac{1-p}{R}-\epsilon\end{array}$$
for some $\epsilon \ge 0$. Substituting the above into \eqref{eq:gamma2_derivative} yields 
\begin{gather*}
\begin{array}{rl}\dot{\Gamma}_{2} = &\sin(\tilde{\theta})( \frac{2(1-p)(1+d_1\xue{(t)})v}{R}
-\epsilon(1+d_1\xue{(t)})v
\\&-2\operatorname{sign}(\sigma)\frac{(1+d_2\xue{(t)})v}{R})\\
\ge & \sin(\tilde{\theta})( \frac{2(1-p)(1+d_1\xue{(t)})v-2(1+d_2\xue{(t)})v}{R}\xue{)}\ge 0,
\end{array}
\end{gather*}
where the first inequality is a consequence of $\sin \tilde{\theta} \le 0$, $\sigma \ge 0 $, and $\operatorname{sign}(\hat{R}(\hat{s})\xue{)}\in \{-1,0,1\}$, while the second inequality holds if $p$ satisfies
\begin{gather}\label{eq:p_range_for_gamma2}
p \geq \max_{\substack{|d_1(t)|\le \bar{d}_1,\\|d_2(t)|\le\bar{d}_2}}(1- \frac{(1 + d_2(t))}{(1 + d_1(t))}) = 
1- \frac{(1 - \bar{d}_2)}{(1 + \bar{d}_1)}.
\end{gather}
Further, since we require $\tilde{\Gamma}_2 \ge \Gamma_2 \ge 0$, \eqref{eq:R_nominal_minimum} is obtained from $\tilde{\Gamma}_2\ge 0$.
Moreover, to enforce both \eqref{eq:p_range_for_gamma1} and \eqref{eq:p_range_for_gamma2}, $p$ must satisfy \eqref{eq:p_range}.

Finally, robust invariance of the closed set  $\mathcal{S}$ is obtained by Nagumo's invariance principle since we have shown that at all points on its boundary $\partial\mathcal{S}$ points inwards. \qed

\subsection{Proof of Theorem 2}

First, we show that the sliding surface/manifold $\sigma=0$ at the intersection of Regions 1 and 3 in Fig. \ref{fig:invariantset} (when $\tilde{\theta}\ge 0$) is attractive, i.e., $V=\frac{1}{2} \sigma^2$ is a Lyapunov function with $\dot{V}=\sigma \dot{\sigma}<0$ for all $\sigma \neq 0$ and $\dot{V}=0$ when $\sigma=0$ such that $\sigma \to 0$ when the state is near the sliding manifold in the second quadrant of Fig. \ref{fig:invariantset}. To achieve this, we first consider Region 1 ($L_1\ge 0$, $0\le\tilde{\theta}\le \frac{\pi}{2}$, $\sigma\ge 0$), in which
\begin{align*}
\begin{array}{rl}
\sigma\dot{\sigma} =\!\!&\sigma(-\frac{\sin(\tilde{\theta})(1-q)(1 + d_1(t))v}{R}
- \operatorname{sign}(\tilde{\theta}) \sin(\tilde{\theta})
\\
&[
-\frac{\cos(\tilde{\theta})(1 + d_1(t))v}{|\hat{R}| - \tilde{y}} 
+ \operatorname{sign}(\sigma)\frac{(1 + d_2(t))v}{R}
])\\
=\!\!\! & - |\sigma| \frac{|\sin \tilde{\theta}|}{R}[(1\!-\!q\!-\! \frac{R\cos \tilde{\theta}}{|\hat{R}|-\tilde{y}})(1 + d_1(t))v\!+\!(1 + d_2(t))v]\\
<\!\!\! & - |\sigma| \frac{|\sin \tilde{\theta}|}{R}[-q(1 + d_1(t))v\!+\!(1 + d_2(t))v]<0,\end{array}
\end{align*}
where the first inequality holds since it can be deduced that $1-\frac{R\cos \tilde{\theta}}{|\hat{R}|-\tilde{y}} \xue{\ >\ } 0$ with $q\in [0,1)$, $|d_1(t)|\le \bar{d}_1$ and $|d_2(t)|\le \bar{d}_2$, while the second inequality holds if $q$ satisfies
\begin{gather}\label{eq:q1}
q \leq \min_{\substack{|d_1(t)|\le \bar{d}_1,\\|d_2(t)|\le \bar{d}_2}}(\frac{(1 + d_2(t))}{(1 + d_1(t))}) = \frac{(1 - \bar{d}_2)}{(1 + \bar{d}_1)}.
\end{gather}
Further, in Region 3 ($\Gamma_1\ge 0$, $L_1\ge 0$, $0\le\tilde{\theta}\le \frac{\pi}{2}$, $\sigma\le 0$),
\begin{align*}
\begin{array}{rl}
\sigma\dot{\sigma} =\!\!&\sigma(-\frac{\sin(\tilde{\theta})(1-q)(1 + d_1(t))v}{R}
- \operatorname{sign}(\tilde{\theta}) \sin(\tilde{\theta})
\\
&[
-\frac{\cos(\tilde{\theta})(1 + d_1(t))v}{|\hat{R}| - \tilde{y}} 
+ \operatorname{sign}(\sigma)\frac{(1 + d_2(t))v}{R}
])\\
=\!\! & - |\sigma| \frac{|\sin \tilde{\theta}|}{R}[(q\!-\!1\!+\! \frac{R\cos \tilde{\theta}}{|\hat{R}|-\tilde{y}})(1 + d_1(t))v\!+\!(1 + d_2(t))v]\\
<\!\! & - |\sigma| \frac{|\sin \tilde{\theta}|}{R}[(q-1)(1 + d_1(t))v\!+\!(1 + d_2(t))v]<0,\end{array}
\end{align*}
where \xue{the first inequality follows from the observation} that $\frac{R\cos \tilde{\theta}}{|\hat{R}|-\tilde{y}}\ge 0$ and the second inequality holds when $q$ satisfies:
\begin{gather}\label{eq:q2}
q \geq \max_{\substack{|d_1(t)|\le \bar{d}_1,\\|d_2(t)|\le \bar{d}_2}}(1 - \frac{(1 + d_2(t))}{(1 + d_1(t))}) = 1-\frac{(1 - \bar{d}_2)}{(1 + \bar{d}_1)}.
\end{gather}

Note that we only proved the attractiveness of sliding surface/manifold at the intersection of Regions 1 and 3. In fact, it can be shown that the other portion of the manifold at the intersection of Regions 2 and 4 is not attractive. Hence, we need to show that
any initial tracking error will converge to the attractive portion of the sliding manifold at the intersection of Regions 1 and 3, which we provide by considering initial errors from each of the \xue{four} different regions. First, if the initial error begins in Region 1 (cf. Fig. \ref{fig:invariantset}), we have shown above that the sliding manifold is attractive. On the other hand, if the tracking error is initialized in Region 3, the tracking error trajectory could either go to the sliding manifold or enter Region 2. It cannot remain in Region 3 without entering Region 2 since we have shown in Theorem \ref{thm:invariance} that $L_1$ is strictly increasing unless $\tilde{\theta}=0$. Similarly, for any tracking error in Region 2, the tracking error trajectory will enter the region of $0 \le \Gamma_2 \le \tilde{\Gamma}_2$ (within Region 4) since we have shown that $L_2$ is strictly increasing unless $\tilde{\theta}=0$. Finally, from there, it will enter Region 1 since $\dot{\Gamma}_2 \ge 0$ in that region unless $\tilde{\theta}=0$, where it will eventually enter the sliding manifold since it is attractive in Region 1.

To conclude that the tracking error converges to zero, we further 
argue that any trajectory starting from the attractive portion of the sliding manifold (when $\tilde{\theta}\ge 0$) will not only stay on the manifold but will flow to the origin where $(\tilde{y},\tilde{\theta})=(0,0)$. The fact that it stays on the sliding manifold is a result of $V = \frac{1}{2} \sigma^2$ (with $\dot{V}=\sigma\dot{\sigma}$ being non-increasing). For the convergence to the origin, we again leverage the proof from Theorem \ref{thm:invariance} that $L_1$ is strictly increasing until $\tilde{\theta}=0$ on the sliding manifold, which is where the origin is. \qed

\fontsize{9}{9}\selectfont
\bibliography{ifacconf}             
\end{document}